\documentclass[twocolumn,floatfix,pra,showpacs]{revtex4-1}

\usepackage{graphicx}
\usepackage{amssymb}
\usepackage{amsmath}
\usepackage{color}
\usepackage{psfrag}
\usepackage{epsfig}
\usepackage{bbm}
\usepackage{bm}
\usepackage{simplewick} 

\newcommand{\ket}[1]{| #1 \rangle}

\newcommand{\rb}[1]{\left( #1 \right)}

\newcommand{\ew}[1]{\langle #1 \rangle}
\newcommand{\beq}{\begin{eqnarray}}
\newcommand{\eeq}{\end{eqnarray}}

\newcommand{\op}[2]{| #1 \rangle \langle #2 |}

\newcommand{\eq}[1]{Eq.~(\ref{#1})}
\newcommand{\fig}[1]{Fig.~\ref{#1}}

\newcommand{\kett}[1]{| #1 \rangle\!\rangle }
\newcommand{\braa}[1]{\langle\!\langle #1|}
\newcommand{\eww}[1]{\langle\! \langle #1\rangle\! \rangle}
\newcommand{\opp}[2]{| #1 \rangle\! \rangle\langle\! \langle #2 |}
\begin{document}
\title{Self-consistent electron counting statistics}
\author{Clive Emary}
\affiliation{
  Institut f\"ur Theoretische Physik,
  Hardenbergstr. 36,
  TU Berlin,
  D-10623 Berlin,
  Germany
}

\date{\today}
\begin{abstract}
We develop a self-consistent version of perturbation theory in Liouville space which seeks to combine the advantages of master equation approaches in quantum transport with the nonperturbative features that a self-consistent treatment brings.  We describe how counting fields may be included in a self-consistent manner in this formalism such that the full counting statistics can be calculated.  NonMarkovian effects are also incorporated.  Several different self-consistent approximations are introduced and we discuss their relative strengths with a simple example.
\end{abstract}
\maketitle

\section{Introduction}

Master equation (ME) techniques play an important role in quantum transport --- not only in describing state-of-the-art experiments e.g. \cite{Sukh07,fri07,shi09,fli09,nad10}, but also in contributing to the theoretical development of the field; topics such as full counting statistics (FCS), both at zero \cite{bag03} and finite \cite{EMAB07} frequency; quantum coherent phenomena \cite{bra05}; and nonMarkovian (NM) effects \cite{bra06,fli08} have all been investigated with master equations.
The central object in such approaches is the Liouvillian or kernel, which generates the evolution of the reduced system density matrix in time and includes not only the internal dynamics of the system but also the effects of the coupling to the leads and to any further environmental degrees of freedom.
ME Liouvillians can be derived in a number of different ways, from the standard textbook approaches, e.g. \cite{coh92,breu02}, to more transport-specific ones such as the wavefunction approach of Gurvitz \cite{gur96}, and the real-time diagrams \cite{sch94,kon96,thi03,thi05}. A reformulation of this latter approach, the Liouvillian perturbation theory (LPT)\cite{sch08, lei08,ce09}, will form the basis of the current work.

Despite their many advantages, standard ME approaches are restricted to the regime in which system and environment are but weakly coupled, with the Liouvillian obtained as a perturbation series in the corresponding coupling strength.  There is much interest in going beyond this situation and to develop nonperturbative approaches for the description of quantum transport beyond the ME.  Some  recent examples include equations of motion \cite{ped05}, renormalisation group \cite{kor07} and self-consistent \cite{cui06,esp10} techniques.

In this paper we develop a {\em self-consistent} (SC) version of the LPT, which seeks to combine the advantages of MEs with a nonperturbative
treatment of the system-lead coupling such that effects such as, for example, level-broadening and higher-order tunnel processes are included.  The method is self-consistent in the sense that system self-energies are calculated not with free propagators, but rather with ones that include the self-energy itself.
We describe how counting fields may be included in a self-consistent manner in this formalism, which means that we are not restricted to the calculation of stationary current only (as in, e.g. \cite{cui06}), but can, in principle, access the complete FCS of the system.  NonMarkovian (NM) effects are also included in our theory from the outset.

Within this framework, we introduce several different SC approximations.  The simplest of these is just a SC version of the familiar sequential-tunneling kernel. This approximation is equivalent to the SC Born approximation \cite{bru04}, appearing in a setting similar to that of Ref.~\cite{cui06}.
By considering the single-resonant level (SRL) model as a test case, we see that although this type of kernel can provide an excellent description of the stationary current even for large couplings ($\Gamma/k_BT \gg 1$) it gives inaccurate results for the shotnoise, and is unsuitable for the calculation of counting statistics in general.
The problems of this kernel are remedied to some extent by the other two kernels described here, which contain not only sequential-type diagrams, but also cotunneling ones.  With these kernels excellent results are obtainable for both current and shotnoise beyond weak coupling.

This work is based on the exposition of the LPT given in Ref.~\cite{ce09}, hereafter referred to as I.  To save needless repetition, the reader will be this paper for many of the details of the original LPT and the notation employed here.

\section{Transport Model \label{secTM}}
We begin by specifying our general transport model with the Hamiltonian 
$
  H = H_\mathrm{res} + H_\mathrm{S} + V
$
composed of reservoir, system, and interaction parts.  In its diagonal basis the system part  reads
$ H_\mathrm{S} = \sum_a E_a \op{a}{a}$ with $\ket{a}$ a many-body state of $N_a$ electrons.
We assume a noninteracting reservoir Hamiltonian
$
  H_\mathrm{res} = \sum_{k,\alpha} (\omega_{k\alpha} +\mu_\alpha)
  a^\dag_{k\alpha} a_{k\alpha}
$, with lead index $\alpha$ that includes spin and any other relevant quantum numbers; $\omega_{k\alpha}$ is the energy of the $k$th mode in lead $\alpha$, $a_{k\alpha}$ a lead annihilation operator, and with $\mu_\alpha$  the chemical potential of lead $\alpha$.
In equilibrium, the reservoir electrons are distributed according to the Fermi function
$
   f(\omega) = 1/\rb{e^{\omega/k_B T} +1}
$,
which, since we assume a uniform temperature, is the same for all reservoirs.
Coupling between system and reservoirs is described by the single-particle tunneling Hamiltonian
\beq
  V =  \sum_{k \alpha m} 
  t_{k \alpha m} a^\dag_{k\alpha} d_m 
  + t^*_{k \alpha m} d^\dag_m a_{k\alpha}
  \label{V1}
  ,
\eeq
where $d_m$ is the annihilation operator for single-particle level $m$ in the system, and $t_{k \alpha m}$ is a tunneling amplitude.

\section{Self-consistent Liouvillian perturbation theory \label{secSCLPT}}

For completeness we give a brief review of the essential elements of the standard Liouvillian perturbation theory --- for more details, the reader is referred I as well as to \cite{sch08,lei08}.
Our starting point is to write down the von Neumann equation for the evolution of the total system-plus-reservoirs density matrix:
\beq
  \dot{\rho}(t) = -i \left[H,\rho(t)\right] =  {\cal L} \rho(t).
  \label{rhodot}
\eeq
This defines the Liouvillian super-operator $ {\cal L} = -i\left[H,\bullet~\right]$, which, in accordance with the decomposition of the Hamiltonian, consists of three parts:
$
  {\cal L} = {\cal L}_\mathrm{res} + {\cal L}_\mathrm{S} + {\cal L}_V
$
with ${\cal L}_\mathrm{res}= -i\left[H_\mathrm{res},\bullet~\right]$,  ${\cal L}_\mathrm{S}=-i\left[H_\mathrm{S},\bullet~\right]$, and 
$ {\cal L}_V = -i\left[V,\bullet~\right]$.
In terms of the Laplace-transformed total density matrix
$
  \rho(z)= \int_{0}^\infty dt e^{-z t} \rho(t)
$,
the solution of \eq{rhodot} is
\beq
  \rho(z) = \frac{1}{z-{\cal L}} \rho(0)
  .
  \label{rhozfull}
\eeq
Assuming a factorizable initial density matrix $\rho(0)$ with reservoirs in equilibrium and system in state $\rho_\mathrm{S}(0)$, tracing over reservoir degrees of freedom results in the following reduced density matrix of the system:
\beq
  \rho_\mathrm{S}(z) 
  = \mathrm{Tr}_\mathrm{res} 
  \left\{
     \rho(z) 
  \right\}
  = \frac{1}{z- {\cal W}(z) } \rho_\mathrm{S}(0)
  \label{rhoSLeff}
  ,
\eeq
with the effective system Liouvillian ${\cal W} (z) = {\cal L}_\mathrm{S} + \Sigma(z)$.  The self-energy $\Sigma(z)$ arises from the coupling with the leads and will be kept in its $z$-dependent nonMarkovian form.

Equation (\ref{rhoSLeff}) is exact if kernel ${\cal W} (z)$ is.  In practice, however, the self-energy contains an infinite sum of terms and must be approximated in some way.
In the straightforward perturbative approach, the memory kernel is calculated as the series $\Sigma(z) = \sum_{n}\Sigma^{(n)}(z)$, with $n$ the number of tunnel vertices in the term, and then approximated by simple truncation at a certain value of $n$.
In previous works, expansion has been considered up to fourth order, such that the self-energy is approximated as
\beq
  \Sigma(z) \approx \Sigma^{(2)}(z)+ \Sigma^{(4)}(z)
  ,
\eeq
with the first term describing sequential tunneling and the second, cotunneling.

In describing the constituent terms of the self-energy and the various modifications to follow, it is useful to employ a diagrammatic representation.  In this representation,  we write the sequential self-energy term as
\beq
  \Sigma^{(2)}(z) &=& 
  \contraction{}{G}{\underset{z}{-}}{G}
  G \underset{z}{-} G. 
  \label{SIG2LPT}
\eeq
The translation of this diagram into an analytical expression is discussed in Appendix \ref{appDIAG}.  Here it suffices to note that the $G$s represent tunnel vertices (of which there are two at sequential order) and these are connected by a single central line that corresponds to the free propagation of the system (i.e. under the action of free system Liouvillian ${\cal L}_\mathrm{S}$).  The overline connecting the tunnel vertices denotes a bath contraction between them.  Such diagrams are similar to those of Ref.~\cite{lei08}, but simplified such that we leave the indices implicit here.

At fourth order in the perturbative expansion of the self-energy there are two diagrams,  each with four tunnel vertices and three free propagators, but with differing patterns of contraction.  In our diagrammatic notation, we write
\beq
  \Sigma^{(4)}(z) =  
  \contraction[2ex]{}{G} {\underset{z}{-} G \underset{z}{-} G \underset{z}{-}} {G}
  \contraction{G \underset{z}{-}} {G} {\underset{z}{-}} {G} 
  G \underset{z}{-} G \underset{z}{-} G \underset{z}{-} G
  +
  \contraction {}{G} {\underset{z}{-} G \underset{z}{-}} {G}
  \contraction[2ex]{G\underset{z}{-}} {G} {\underset{z}{-} G \underset{z}{-}} {G}
  G \underset{z}{-} G \underset{z}{-} G \underset{z}{-} G
  \label{SIG4LPT}
  ,
\eeq 
with the first term the ``direct'' contribution and the second, the ``exchange''.
We have left the $z$-dependence of the propagators explicit here, as this will be important for the calculation of NM effects.

\subsection{Self-consistent kernels}

The main idea behind this paper is to take the kernels obtained in LPT and replace in them the free system propagators with their effective counterparts that already include the effects of the reservoirs.
The simplest SC kernel that one can come up with in this manner, which we denote $\Sigma^{(a)}(z)$, is simply the SC equivalent of the sequential kernel of \eq{SIG2LPT}:
\beq
  \Sigma^{(a)}(z) =
  \contraction{}{G}{ \underset{z}{\eqcirc}}{G}
  G \underset{z}{\eqcirc} G
  \label{SIGa}
  ,
\eeq
in which the propagator ``$ \eqcirc$'' contains the effective system Liouvillian
${\cal W}^{(a)}(z) = {\cal L}_\mathrm{S}+\Sigma^{(a)}(z)$ rather than just free Liouvillian ${\cal L}_\mathrm{S}$. 
Expansion of this propagator shows that this self-energy reproduces the sequential term exactly, and provides an approximation to the direct cotunneling term and all higher diagrams in which no contraction lines cross.
This kernel is equivalent to a SC Born approximation.

The second SC kernel that we propose here reads
\beq
  \Sigma^{(b)}(z) =
  \contraction{}{G}{\underset{z}{\eqcirc}}{G}
  G \underset{z}{\eqcirc} G
  +
  \contraction {}{G} {\underset{z}{\eqcirc} G \underset{z}{\eqcirc}} {G}
  \contraction[2ex]{G\underset{z}{\eqcirc}} {G} {\underset{z}{\eqcirc} G \underset{z}{\eqcirc}} {G}
  G \underset{z}{\eqcirc} G \underset{z}{\eqcirc} G\underset{z}{\eqcirc} G
  \label{SIGb}
  ,
\eeq
where the propagator here contains Liouvillian
${\cal W}^{(b)}(z) = {\cal L}_\mathrm{S}+\Sigma^{(b)}(z)$. 
This kernel explicitly includes not only a sequential-type diagram but also that of the fourth-order exchange term.
Expansion shows that $\Sigma^{(b)}(z)$ contains both exchange and direct cotunneling contributions at fourth order --- the exchange term exactly, and an approximation to the direct term from the SCBorn part. At higher orders this kernel includes a far larger class of diagrams than does $\Sigma^{(a)}(z)$, both with and without crossings.   As we will see, this represents a significant advance over the SCBorn kernel, $\Sigma^{(a)}(z)$, not least because there is a large degree of cancellation between direct and exchange contributions.
Lastly we also introduce the SC kernel
\beq
  \Sigma^{(c)}(z) &=&
  \contraction{}{G}{\underset{z}{-}}{G}
  G \underset{z}{-} G
  +
  \contraction[2ex]{}{G} {\underset{z}{-} G \underset{z}{\eqcirc} G \underset{z}{\eqcirc}} {G}
  \contraction{G \underset{z}{-}} {G} {\underset{z}{\eqcirc}} {G} 
  G \underset{z}{-} G \underset{z}{\eqcirc} G \underset{z}{\eqcirc} G
  \nonumber\\
  &&+
  \contraction {}{G} {\underset{z}{\eqcirc} G \underset{z}{\eqcirc}} {G}
  \contraction[2ex]{G\underset{z}{-}} {G} {\underset{z}{\eqcirc} G \underset{z}{\eqcirc}} {G}
  G \underset{z}{-} G \underset{z}{\eqcirc} G\underset{z}{\eqcirc} G
  \label{SIGc}
  ,
\eeq
which explicitly contains diagrams of both cotunneling types, as well as a mix of free and effective propagators.  This kernel has the advantage that the sequential contribution is parameterised independently of higher order terms. Expansion shows that this kernel is exact up to fourth order.
These three kernels all share the property that, upon expansion, they give a sum of diagrams with topologies the same as a subclass of diagrams of the exact kernel with no overcounting.

Translated into an equation for the effective system Liouvillian, \eq{SIGa} gives the following functional form
\beq
  {\cal W}^{(a)}(z) = {\cal L}_\mathrm{S} +\int d \omega_1 F^{(a)}({\cal W}^{(a)}(z),\omega_1 )
  \label{SIGFa}
  ,
\eeq
where $ F^{(a)}$ is some function (the form of which is irrelevant for the current discussion, but can be found in Appendix \ref{appINTEGRALS}), and where we have made explicit the integration over bath frequency $\omega_1$.  The corresponding functional form of \eq{SIGb} is the double integral
\beq
  {\cal W}^{(b)}(z) = {\cal L}_\mathrm{S} +\int d \omega_1 \int d \omega_2 F^{(b)}(  {\cal W}^{(b)}(z),\omega_1 ,\omega_2)
  \label{SIGFb}
  ,
\eeq
with a similar equations for $  {\cal W}^{(c)}(z)$.
In all cases, we can make progress with these integral equations by introducing the eigendecomposition of ${\cal W}$ on the righthandside (Appendix \ref{appMATRIX}).  The integrals can then be evaluated analytically as functions of the eigenvalues of ${\cal W}$ (Appendix \ref{appINTEGRALS}) and the resulting matrix equations can then be solved numerically by iteration.

\section{Counting Statistics  \label{secCS}}
To calculate the current and its fluctuations we will employ the FCS formalism \cite{lev93,yul99} as appropriate for master equation calculations \cite{bag03,fli04,jau05,flindtthesis,fli08,bra09}.
The central object in this formalism is the counting-field-resolved Liouvillian ${\cal W}(\chi;z) = {\cal L}_\mathrm{S} + \Sigma(\chi;z)$,
in which processes that transfer $n$ electrons to and from the leads are associated with counting factors $e^{\pm i n \chi_\alpha}$, where $\chi_\alpha$ is the counting field and $\alpha$ labels the lead in which counting takes place. Once in possession of the $\chi$-resolved Liouvillian, all zero-frequency cumulants of the current can, in principle, be calculated.
Given the existence of this well-developed formalism, the outstanding question for us to answer here is how to include the counting fields into our self-consistent kernels.

Achieving this requires two modifications to definitions \eq{SIGa}, \eq{SIGb}, and \eq{SIGc}.  Firstly, as in the LPT calculation of I, the bath contractions are modified to contain counting fields:
\beq
   \gamma_{21}^{p_2 p_1} \to
   \gamma_{21}^{p_2p_1}(\chi)=\gamma_{21}^{p_2p_1}
   e^{
     i s_{\alpha_1} \xi_1 \frac{1}{2} \rb{p_1-p_2} \chi_{\alpha_1}
   }
   \label{gamchi}
   ,
\eeq
with $\chi_{\alpha_1}$, the counting field of lead $\alpha_1$;
$s_\alpha=\pm 1$, a factor given by the sign-convention for current flow in lead $\alpha$; $p_1$ and $p_2$, Keldysh indices; and $\xi_1=\pm 1$, an index indicating whether an electron enters or leaves the system at vertex $1$ (see Appendix \ref{appDIAG} and I for a full explanation of these symbols).
Secondly, the SC self-energies are not to be evaluated using propagator ``$\underset{z}{\eqcirc}$'', but rather its $\chi$-dependent counterpart ``$\underset{\chi;z}{\eqcirc}$'', which describes propagation under the action of $\chi$-dependent kernel ${\cal W}(\chi;z)$.
The equation for the $\chi$-dependent SC kernel of type $(a)$ thus reads
\beq
  \Sigma^{(a)}(\chi;z) = 
  \contraction{}{G}{\underset{\chi;z}{\eqcirc}}{G}
  G \underset{\chi;z}{\eqcirc} G ~q_2(\chi)  
  \label{SIGchia} 
  ,
\eeq
where the propagator on the righthandside is evaluated with the full nonMarkovian $\chi$-dependent effective Liouvillian ${\cal W}^{(a)}(\chi;z) = {\cal L}_\mathrm{S}+ \Sigma^{(a)}(\chi;z)$ and where
\beq
  q_2(\chi) &=& 
  e^{
     i s_{\alpha_1} \xi_1 \frac{1}{2}\rb{p_1-p_2}\chi_{\alpha_1}
  }
  \label{q2}
  ,
\eeq 
is the counting-field factor arising from the single bath contraction at sequential order.
Similarly, for the other two schemes we have
\beq
  \Sigma^{(b)}(\chi;z) = 
  \contraction{}{G}{\underset{\chi;z}{\eqcirc}}{G}
  G \underset{\chi;z}{\eqcirc} G ~q_2(\chi)
  +
  \contraction {}{G} {\underset{\chi;z}{\eqcirc} G \underset{\chi;z}{\eqcirc}} {G}
  \contraction[2ex]{G \underset{\chi;z}{\eqcirc}} {G} {\underset{\chi;z}{\eqcirc} G \underset{z}{\eqcirc}} {G}
  G \underset{\chi;z}{\eqcirc} G \underset{\chi;z}{\eqcirc} G \underset{\chi;z}{\eqcirc} G ~ q_{4X}(\chi)
  ,
  \nonumber\\
  \label{SIGchib}
\eeq
and
\beq
  \Sigma^{(c)}(\chi;z) &=& 
  \contraction{}{G}{\underset{\chi;z}{-}}{G}
  G \underset{z}{-} G ~q_2(\chi)  
  +
  \contraction[2ex]{}{G} {\underset{z}{-} G \underset{\chi;z}{\eqcirc} G \underset{\chi;z}{\eqcirc}} {G}
  \contraction{G \underset{z}{-}} {G} {\underset{\chi;z}{\eqcirc}} {G} 
  G \underset{z}{-} G \underset{\chi;z}{\eqcirc} G \underset{\chi;z}{\eqcirc} G
   ~ q_{4D}(\chi)
   \nonumber\\
   &&
  +
  \contraction {}{G} {\underset{\chi;z}{\eqcirc} G \underset{\chi;z}{\eqcirc}} {G}
  \contraction[2ex]{G \underset{z}{-}} {G} {\underset{\chi;z}{\eqcirc} G \underset{z}{\eqcirc}} {G}
  G \underset{z}{-} G \underset{\chi;z}{\eqcirc} G \underset{\chi;z}{\eqcirc} G ~ q_{4X}(\chi)
  \label{SIGchic} 
  ,
\eeq
with counting-field factors
\beq
  q_{4X}(\chi) &=& 
  e^{
   i s_{\alpha_1} \xi_1 \frac{1}{2} \rb{p_1-p_3} \chi_{\alpha_1}
  }
  e^{
   i s_{\alpha_2} \xi_2 \frac{1}{2}\rb{p_2-p_4} \chi_{\alpha_2}
  }  
  \nonumber\\
  q_{4D}(\chi) &=& 
  e^{
   i s_{\alpha_1} \xi_1 \frac{1}{2} \rb{p_1-p_4} \chi_{\alpha_1}
  }
  e^{
   i s_{\alpha_2} \xi_2 \frac{1}{2}\rb{p_2-p_3} \chi_{\alpha_2}
  } 
  ,
\eeq
from the two bath contractions at fourth order. 

The justification for including the $\chi$-dependent self energies $\Sigma(\chi;z)$ on the righthandside propagators, instead of e.g. $\Sigma(\chi=0;z)$,
comes from expanding these forms and comparing with an exact $\chi$-dependent LPT expansion.  Moreover, including $\chi$ in this fashion ensures charge conservation as discussed in the section \ref{secCC}.

\subsection{Current correlations \label{SUBSECcorrel}}

Solving the self-consistent equations \eq{SIGchia}, \eq{SIGchib}, or \eq{SIGchic} for the full $\chi$-dependent $\Sigma(z;\chi)$ as a function of $z$ and $\chi$ is most likely too demanding to be achieved in practice.  Nevertheless, these equations may be used to generate the various quantities required to calculate current correlation functions.
Here, we calculate the zero-frequency current correlation functions using the nonMarkovian ``current block'' formalism of Refs.~\cite{fli08,flindtthesis,ce09}, which does not require the full $\Sigma(z;\chi)$, but rather its various derivatives evaluated at $z=\chi=0$.

We  work in a representation in which the elements of the system density matrix are arranged into a vector, such that Liouvillian ${\cal W}(z)$ can be written as a matrix.
The stationary state of the system, which we write as $\rho^\mathrm{stat}_\mathrm{S}=\kett{\psi_0}$ (see Appendix \ref{appMATRIX}), can be obtained as the right null-vector of the zero-frequency limit of ${\cal W}(z=0)$:
${\cal W}(0) \kett{\psi_0}=0$,
and this we assume to be unique. 
Similarly, the left null-vector defined via
$
 = \braa{\psi_0} {\cal W}(0) =0
$ is the system ``trace'' vector \cite{jau05}.  We further define the stationary state ``expectation value'' $\eww{\bullet} = \eww{\psi_0 | \bullet |\psi_0}$, and the projectors ${\cal P} = \opp{\psi_0}{\psi_0}$ and ${\cal Q} = \mathbbm{1}-{\cal P}$.
We then define
\beq
  {\cal J}(\chi,\epsilon) = {\cal W}(\chi,z=0-i\epsilon)-{\cal W}(\chi=0,z=0)
  ,
\eeq
with the derivatives
$
  {\cal J}'= \left. \partial_{\chi} {\cal J}\right|_{\chi,\epsilon \to 0}
$,
$
  \dot{\cal J} = \left. \partial_\epsilon {\cal J}\right|_{\chi,\epsilon \to 0}
$,
and analogously for higher-orders. We also require the zero-frequency pseudo-inverse
\beq
  {\cal R} =  \lim_{\epsilon\to 0}{\cal Q} 
  \frac{1}{i \epsilon + {\cal W}(0, -i\epsilon)} {\cal Q}
  \label{pseudoR}
  .
\eeq
The stationary average current can then simply be written as
\beq
  \ew{I} &=& \eww{{\cal J}'} 
  \label{avI}
  ,
\eeq
with the zero-frequency shotnoise, $S$, given by
\beq
  i^2 S &=& 
  \eww{{\cal J}''}
  -2 \eww{{\cal J}' {\cal R} {\cal J}'}
  -2 \ew{I} 
  \rb{
    \eww{\dot{{\cal J}}'}
    - \eww{{\cal J}' {\cal R} \dot{{\cal J}}}
    }
  \label{S}
  \nonumber\\
  .
\eeq
The final term in \eq{S} is the nonMarkovian correction.   Expressions for higher current cumulants can be found in Refs.~\cite{flindtthesis}, and obtained recursively \cite{fli08}, but we focus here on the average current and shotnoise.

Equations for the blocks required in the above expressions can be obtained by differentiating the definitions of the kernels \eq{SIGchia}, \eq{SIGchib} or \eq{SIGchic}, and setting $z,\chi \to 0$.  For the purpose of exposition, we consider just the $(a)$-type kernel here; the equations for the $(b)$-kernel are given in appendix \ref{appSIGB}.   The first derivatives of  \eq{SIGchia} yield the equations
\beq
  {\cal J}' &=& 
  \contraction{}{G}{\eqcirc}{G}
  G \eqcirc G ~q_2'
  +
  \contraction{}{G}{\eqcirc{\cal J}'\eqcirc}{G}
  G \eqcirc{\cal J}'\eqcirc G 
  \label{J'}\\
  \dot{{\cal J}} &=& 
  \contraction{}{G}{\eqcirc\rb{i\mathbbm{1}+\dot{{\cal J}}}\eqcirc}{G}
  G \eqcirc\rb{i \mathbbm{1}+\dot{{\cal J}}}\eqcirc G,
  \label{Jdot}
\eeq
and the required second-order derivatives give
\beq
  \eww{{\cal J}''} &=& 
  \eww{
	  \contraction{}{G}{\eqcirc}{G}
	  G \eqcirc G 
	}~q_2''
  +~
  2 
  \eww{  
	  \contraction{}{G}{\eqcirc{\cal J}'\eqcirc}{G}
	  G \eqcirc{\cal J}'\eqcirc G 
	}~q_2'
  \label{J''}\\
  \eww{\dot{{\cal J}}'} &=& 
  \eww{
	  \contraction{}{G}{\eqcirc\rb{i\mathbbm{1}+\dot{{\cal J}}}\eqcirc}{G}
	  G \eqcirc\rb{i \mathbbm{1}+\dot{{\cal J}}}\eqcirc G 
	  } ~q_2'
	  \label{J'dot}
\eeq
In the latter case, we give expressions for the stationary expectation values of the blocks, rather than the blocks themselves, since these are simpler with many terms giving zero in the expectation value thanks to the ``leftmost-$G$ rule'' \cite{leftmostG}.  
The evaluation of diagrams of the form 
$
  \contraction{}{G}{\eqcirc M\eqcirc}{G}
  G \eqcirc M\eqcirc G 
$ is discussed in Appendix \ref{appINTEGRALS}. Equations (\ref{J'}) and (\ref{Jdot})
each represent a set of linear equations for the matrix elements of ${\cal J}'$ and $\dot{{\cal J}}$ respectively, which can easily be solved and the results substituted into Eqs.(\ref{J''}) and (\ref{J'dot}).

Given the above formal developments, it is perhaps useful to give a brief overview of its practical application. Let us refer to the procedure for the $(a)$-type kernel.
Firstly, all the basic elements of LPT are constructed as matrices, e.g. the free Liouvillian, the various tunnel operators, etc. Next the stationary kernel $\Sigma^{(a)}(0;0)$ is determined using \eq{SIGa}. This equation is solved iteratively with, for example, the standard LPT kernel as starting point.  Once $\Sigma^{(a)}(0;0)$ is known, the stationary state $\rho^\mathrm{stat}_\mathrm{S}$ and pseudo-inverse ${\cal R}$ can easily be found. Equations (\ref{J'}) and (\ref{Jdot}) are then constructed, solved for ${\cal J}'$ and $\dot{{\cal J}}$ and the results being fed into Eqs. (\ref{J''}) and (\ref{J'dot}) to obtain  $\eww{{\cal J''}}$ and $\eww{\dot{{\cal J}}'}$. These various elements are then combined in accordance with the expressions \eq{avI} and \eq{S} to give the average current and shotnoise.

\section{Charge conservation \label{secCC}}

Before applying the method, we first demonstrate that it respects charge conservation. With charge conserved, a difference in instantaneous currents gives rise to a change in the accumulated system charge $\hat{Q}=\sum_a N_a\op{a}{a}$ with $N_a$ the number of electrons in state $a$ ($e=1$).  In operator terms we have
$
   \sum_\alpha s_\alpha \hat{I}_\alpha = -d\hat{Q}/dt
$,
where currents $s_\alpha I_\alpha$ are defined positive for electron flow into reservoir $\alpha$.  
Taking the stationary expectation value gives immediately the relation
$\sum_\alpha s_\alpha \ew{I_\alpha}=0$
for the average currents.  
In terms of the jump super-operators of section \ref{SUBSECcorrel}, this relation reads 
\beq
  \sum_\alpha s_\alpha \ew{I_\alpha} =  
  \sum_\alpha s_\alpha \eww{ {\cal J}'_\alpha}
  = 0
  \label{meanconserve}
\eeq
with block ${\cal J}'_\alpha$ is defined as in \eq{J'}, but with the lead index of the $\chi$-derivative made explicit.
Similarly, for the shotnoises, charge conservation implies \cite{bla00}
\beq
  \sum_{\alpha \beta} s_\alpha s_\beta S_{\alpha \beta}(0)= 0
  ,
\eeq
with the zero-frequency shotnoise correlator $S_{\alpha \beta}(0)$ between leads $\alpha$ and $\beta$.  Using the lead-specific analogue of \eq{S}, in terms of jump operators this relation reads 
\beq
   \sum_{\alpha \beta} s_\alpha s_\beta \eww{
    J_{\alpha \beta}'' - 2J_\alpha ' {\cal R}  J_\beta '
  }=0
  ,
\eeq
where the application of \eq{meanconserve} makes the nonMarkovian contributions cancel under the sum.

We show that our approach obeys these relations by extending the rate-equation work of Ref.~\cite{bag03} to the SC kernels studied here.  Considering the $\Sigma^{(a)}$ SC-sequential kernel as an example, explicit 
multiplication alows us to calculate various commutators involving system charge operator $\hat{Q}$.   Firstly, the commutator of $\hat{Q}$ with the full $\chi$-dependent kernel of \eq{SIGchia} reads 
\beq
  \left[
    {\cal W}(\chi),\hat{Q}
  \right] = 
  \sum_\alpha s_\alpha
  \contraction{}{G}{\underset{\chi}{\eqcirc}}{G}
  G \underset{\chi}{\eqcirc}G ~q'_\alpha(\chi)
\eeq
(for brevity, we write here ${\cal W}(\chi)={\cal W}^{(a)}(\chi;z=0)$).
At $\chi=0$, this relation yields
\beq
  \left[
    {\cal W}(0),\hat{Q}
  \right] = 
  \sum_\alpha s_\alpha
  \contraction{}{G}{\underset{~}{\eqcirc}}{G}
  G \underset{~}{\eqcirc}G ~q'_\alpha(0)
  \label{SCWn1}
\eeq
Differentiating with respect to $i\chi_\beta$ , setting $\chi_\beta \to 0$ and summing over $\beta$ with lead-factors $s_\beta$ we also obtain
\beq
 \left[ \left[
    {\cal W}(0),\hat{Q}
  \right],\hat{Q}
  \right] 
  +
  \left[ 
    \sum_\beta s_\beta
    \contraction{}{G}{\eqcirc{\cal J}_\beta'\eqcirc}{G}
    G \eqcirc{\cal J}_\beta'\eqcirc G 
    ,\hat{Q}
  \right]
  \nonumber\\
  =
  \sum_\alpha s_\alpha s_\beta
  \left\{
    \contraction{}{G}{\underset{~}{\eqcirc}}{G}
    G \underset{~}{\eqcirc}G ~q''_{\alpha\beta}(0)
    +
    \contraction{}{G}{\eqcirc{\cal J}_\beta'\eqcirc}{G}
    G \eqcirc{\cal J}_\beta'\eqcirc G 
    ~q'_\alpha
  \right\}
  \label{SCWnn1}
  .
\eeq
An identical calculation can be performed without the $\chi$-dependence of the propagator in ${\cal W}(\chi)$ which yields
\beq
  \left[
    \contraction{}{G}{\underset{\chi=0}{\eqcirc}}{G}
    G \underset{\chi=0}{\eqcirc}G ~q(\chi)
    ,\hat{Q}
  \right] = 
  \sum_\alpha s_\alpha
  \contraction{}{G}{\underset{\chi=0}{\eqcirc}}{G}
  G \underset{\chi=0}{\eqcirc}G ~q'_\alpha(\chi)
  .
\eeq
This recovers \eq{SCWn1} and gives additionally
\beq
 \left[ \left[
    {\cal W}(0),\hat{Q}
  \right],\hat{Q}
  \right] 
  = 
  \sum_\alpha s_\alpha s_\beta
    \contraction{}{G}{\underset{0}{\eqcirc}}{G}
    G \underset{0}{\eqcirc}G ~q''_{\alpha\beta}(0)
  \label{SCWnn2}
  .
\eeq
Comparison of \eq{SCWnn1} and \eq{SCWnn2} then yields
\beq
  \left[ 
    \sum_\beta s_\beta
    \contraction{}{G}{\eqcirc{\cal J}_\beta'\eqcirc}{G}
    G \eqcirc{\cal J}_\beta'\eqcirc G 
    ,\hat{Q}
  \right]
  = 
  \sum_\alpha s_\alpha s_\beta
    \contraction{}{G}{\eqcirc{\cal J}_\beta'\eqcirc}{G}
    G \eqcirc{\cal J}_\beta'\eqcirc G 
    ~q'_\alpha
    \nonumber
  \label{SCWn3}
  .
\eeq

Equation \ref{SCWn1} allows us to write 
\beq
  \sum_\alpha s_\alpha \eww{{\cal J}_\alpha'}
  &=&
  \eww{\left[
    {\cal W}(0),n
  \right]}
  + 
  \sum_\alpha s_\alpha
  \eww{
  \contraction{}{G}{\eqcirc{\cal J}_\alpha'\eqcirc}{G}
  G \eqcirc{\cal J}_\alpha'\eqcirc G 
  }
  \nonumber\\
  &=&0
  ,
\eeq
where the first term vanishes due to the action of ${\cal W}(0)$ in the stationary expectation value, and the second due to  the `` leftmost-$G$ rule'' \cite{leftmostG}. 
Similarly, using \eq{SCWnn1} and \eq{SCWn3}, the sum of shotnoises evaluates as
\begin{widetext}
\beq
  \sum_{\alpha \beta} s_\alpha s_\beta S_{\alpha \beta}(0)
  &=&
  \eww{
  \left[ \left[ {\cal W},\hat{Q}  \right],\hat{Q} \right]
  +2
  \left[ 
    \sum_\beta s_\beta
    \contraction{}{G}{\eqcirc{\cal J}_\beta'\eqcirc}{G}
    G \eqcirc{\cal J}_\beta'\eqcirc G 
    ,\hat{Q}
  \right]
  }
  \nonumber\\
  &&
  -2
  \eww{
    \rb{
      \left[{\cal W},\hat{Q}\right] 
      + 
  \sum_\alpha s_\alpha
  \contraction{}{G}{\eqcirc{\cal J}_\alpha'\eqcirc}{G}
  G \eqcirc{\cal J}_\alpha'\eqcirc G 
    }
      {\cal Q} {\cal W}^{-1} {\cal Q} 
    \rb{   
      \left[{\cal W},\hat{Q}\right]
      + 
  \sum_\alpha s_\alpha
  \contraction{}{G}{\eqcirc{\cal J}_\alpha'\eqcirc}{G}
  G \eqcirc{\cal J}_\alpha'\eqcirc G 
    }
  }
  \nonumber\\
  &=&
  \eww{
   -2 \hat{Q}{\cal P} 
   \sum_\beta s_\beta
    \contraction{}{G}{\eqcirc{\cal J}_\beta'\eqcirc}{G}
    G \eqcirc{\cal J}_\beta'\eqcirc G 
  } = 0
  ,
\eeq
\end{widetext}
once again courtesy of the properties of ${\cal W}(0)$ and the leftmost-$G$ rule.

The mean currents and shotnoises therefore obey the properties demanded of them by charge conservation.  An identical argument can be made for the other self-consistent kernels considered here. It is also relatively straightforward to extend this argument to higher cumulants and to other kernels.

\section{Results \label{secRES}}
As an example application of this formalism, we consider a quantum dot with a single spinless level.  The Hamiltonian reads
$
  H = 
   \varepsilon d^\dagger d
  + H_\mathrm{res}
  + V
$,
with $\varepsilon$ the energy of the level when occupied and with lead and coupling Hamiltonians $H_\mathrm{res}$ and $V$ as in Sec.~\ref{secTM}.  This model acts as an important benchmark for transport calculations, e.g. \cite{zed09}, as interactions are absent and exact solutions are available from, e.g., scattering theory \cite{but92,bla00}. The exact results were summarised in I.

\begin{figure}[tb]
  \psfrag{GeqkBT}{$\Gamma_R=k_BT$}
  \psfrag{Geq3kBT}{$\Gamma_R=3k_BT$}
  \psfrag{Geq5kBT}{$\Gamma_R=5k_BT$}
  \psfrag{eVkBT}{$eV/k_BT$}
  \psfrag{eV}{$eV/k_BT$}
  \psfrag{IkBT}{$\ew{I}/k_BT$}
  \psfrag{error}{$|\delta I|$}
  \begin{center}
  \epsfig{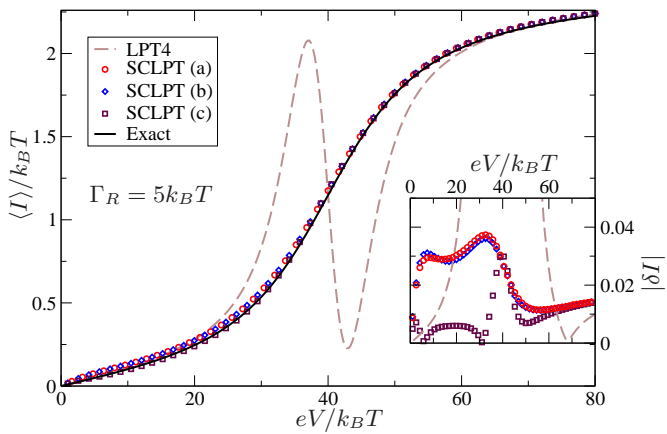}
  \caption{
   Stationary current $\ew{I}$ through the single resonant level as a function of applied bias $eV$ with system-lead coupling $\Gamma_L=\Gamma_R = 5 k_BT$.
   The main panel shows the current itself with results plotted for the three SC kernels discussed here, together with the exact solution and the fourth-order LPT results.  For this coupling, the LPT solution breaks down when dot level is located around resonance with the chemical potentials.  In contrast, all three SC results follow the exact solution closely.  The inset shows the absolute value of the difference $\delta I$ between exact and each approximate solution.
   Further parameters were dot level position $\varepsilon=20 k_B T$, chemical potentials $\mu_L=-\mu_R =  e V/2$, and cut-off $X_C=300 k_B T$.
    \label{figSRLI}
 }
  \end{center}
\end{figure}
In \fig{figSRLI} we show the current calculated with our three SC kernels alongside the exact result and that from straightforward fourth-order LPT.  Results are shown for a system-lead coupling of $\Gamma_R=\Gamma_L = 5 k_B T$.  
As is clear, at this coupling the fourth-order perturbative expansion LPT solution breaks down badly when the dot level is located around resonance with the chemical potentials.  This is in line with the finding of Refs.\cite{thi05,ce09}, which found the fourth-order ME calculation to be reliable across the full bias range for a coupling $\Gamma_R/k_B T\lesssim 1/2$.
In contrast, all three SC kernels give a very good account of the current.  Scrutiny of the inset, which shows the difference between exact an approximate solutions, shows that $\Sigma^{(c)}$ is the most accurate  across the bias range, and that $\Sigma^{(a)}$ gives comparable results to the other kernels despite being much easier to calculate.

\begin{figure}[tb]
  \psfrag{GeqkBT}{$\Gamma_R=k_BT$}
  \psfrag{Geq3kBT}{$\Gamma_R=3k_BT$}
  \psfrag{Geq5kBT}{$\Gamma_R=5k_BT$}
  \psfrag{eVkBT}{$eV/k_BT$}
  \psfrag{SkBT}{$S/k_BT$}
  \begin{center}
  \epsfig{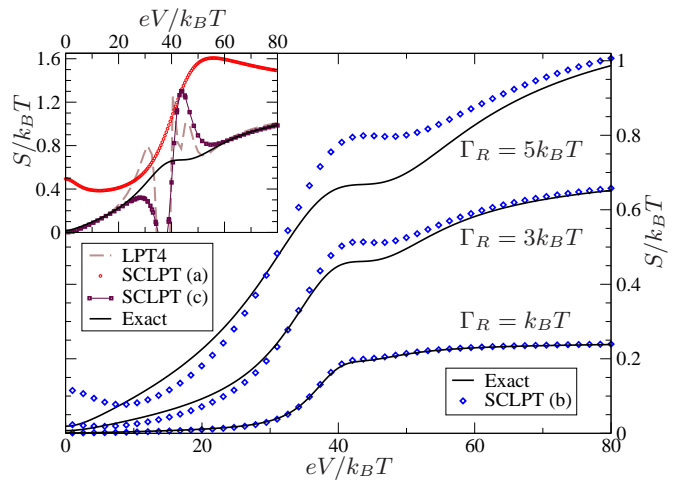}
  \caption{
    Zero-frequency shotnoise $S$ for the single resonant level as a function of bias.  The main panel shows a comparison of the 
    results from the self-consistent $\Sigma^{(b)}$ scheme with the exact results for three different couplings: $\Gamma_R=\Gamma_L=1,3,5 ~k_BT$.
    The inset shows results for fourth-order LPT and the other two SC kernels for a coupling of $\Gamma_R=\Gamma_L=5 k_BT $.  Other parameters as in \fig{figSRLI}.
    \label{figSRLS}
 }
  \end{center}
\end{figure}
Figure \ref{figSRLS} shows results for the shotnoise of this model.  Starting with the inset we see that that the LPT4 results once again break down around resonance.  Moreover, we see that neither $\Sigma^{(a)}$ nor  $\Sigma^{(c)}$ provides a good description across the whole bias range. 
The results for $\Sigma^{(c)}$ break down around resonce in a fashion similar to the LPT solution and this can be attributed to the inclusion of unbroadened propagators in the kernel.  Results at large and small bias for this kernel are nevertheless very good.
Results for kernel $\Sigma^{(a)}$ break down in a different manner; there is no abrupt behaviour around resonance, but the predicted value of the shotnoise is everywhere much higher than the exact solution.
 This difference is significant and remains even when the coupling is reduced.  This error can be traced to the failure to includes diagrams with crossings.  In particular, at fourth order, direct and exchange contributions are of similar magnitude, but with opposite sign.  Failure to include diagrams with crossings therefore leads a large portion of the direct contribution remaining uncancelled, and this leads to the overestimated shotnoise.  This problem is not manifest in the current because, for this model at least, the fourth-order exchange term does not contribute anyway.  We must therefore conclude that this SCBorn-type kernel is unreliable for calculating shotnoise, and by extension higher counting statistics in this intermediate coupling regime.  

The main panel shows results for kernel  $\Sigma^{(b)}$ and we see that this kernel does give a good account of the shotnoise across the bias range.
For higher couplings, the SC results remain qualitatively correct except at low bias, when the system is in the cotunneling regime.  Here the SC solution shows a spurious increase of the the shotnoise with decreasing bias.  This arises from the conflict of requiring both that the sequential contribution vanish in this cotunneling limit, but also be finite to generate the correct direct cotunneling diagram under iteration.  Clearly these two requirements can not be met at the same time, and this leads to a loss of accuracy in this regime.

\section{Conclusions \label{secCONC}}

In this paper we have described a method for calculating counting statistics using self-consistent kernels.  Three different SC kernels were introduced and we have considered current and shotnoise for the single resonant level as an example application.  For this model, all three kernels provide good results for the current, significantly better than could be achieved with straightforward perturbative techniques.  On the other hand, only $\Sigma^{(b)}$ is able to provide good results for the shotnoise, in particular when the dot level is around resonance with the chemical potentials. This is the most interesting regime since simple LPT performs well at both high bias and in the cotunneling regime.

Although the above comparison is made for a non-interacting model, we expect the conclusions to hold broadly true for interacting models in the intermediate coupling regime considered here.
It remains as future work to see how these SC techniques perform in strongly-coupled interacting systems in which genuine nonperturbative effects, such as the Kondo resonance \cite{kondotheory,kondoexp}, come into play.

\begin{acknowledgments}
Work supported by the WE Heraeus foundation and by DFG grant BR 1528/5-1.  I am grateful to T.~Brandes and R.~Aguado for their helpful input.
\end{acknowledgments}

\appendix

\section{Diagrams \label{appDIAG}}

Although we will leave much of the detail of the LPT expansion to I, it behoves us here to discuss those elements relevant to the translation of  our diagrams into the corresponding analytic LPT expressions.  The (nonSC-) LPT sequential diagram of \eq{SIG2LPT} reads
\beq
  \Sigma^{(2)}(z) &=& 
  \contraction{}{G}{\underset{z}{-}}{G}
  G \underset{z}{-} G
  ,
\eeq
and this translates as
\beq
  \Sigma^{(2)}(z) &=& (-1)
  G^{p_2}_2 
  \Omega_\mathrm{S}(z_1)
  G^{p_1}_1 \gamma_{21}^{p_2 p_1}
  \label{sig2translate}
  .
\eeq
Similarly, the fourth-order exchange term has the diagram
\beq
  \Sigma^{(4X)}(z) &=&
  \contraction {}{G} {\underset{z}{-} G \underset{z}{-}} {G}
  \contraction[2ex]{G\underset{z}{-}} {G} {\underset{z}{-} G \underset{z}{-}} {G}
  G \underset{z}{-} G \underset{z}{-} G\underset{z}{-} G
  ,
\eeq
which translates as
\beq
  \Sigma^{(4X)}(z) &=&
  (-1)
  G^{p_4}_4 
  \Omega_\mathrm{S}(z_3)
  G^{p_3}_3
  \Omega_\mathrm{S}(z_2)
  \nonumber\\
  &&~~~
  \times
  G^{p_2}_2 
  \Omega_\mathrm{S}(z_1)
  G^{p_1}_1  
  \gamma_{42}^{p_4 p_2}\gamma_{31}^{p_3 p_1}
  \label{sig4Xtranslate}
  .
\eeq
In these expressions $G^{p}_{1}$ are Liouville-space tunnel vertices which add or remove electrons from the system, with multi-index subscript `$1$'$=(\xi_1,k_1,\alpha_1)$  denoting  direction $\xi=\pm1$, bath momentum $k_1$ and lead ($\alpha=L,R$) indices, and with Keldysh index superscript ($p_i=\pm 1$).  These diagrams include the free system propagator, which translates as
\beq
   \underset{z}{-} \to \Omega_\mathrm{S}(z_m) 
   = \frac{1}{z_m-{\cal L}_\mathrm{S}}
   ,
\eeq
with free system Liouvillian ${\cal L}_\mathrm{S}$ and with the subscript $m$ determined by the position of the propagator in the diagram. The frequencies $z_m$ are given by $z_m = z+\sum_{l=m+1}^n x_l$ with Laplace variable $z$, $n$ the order of the diagram (e.g. $n=2$ for sequential order) and $x_l=-i\xi_l(\omega_l+\mu_{\alpha_l})$,  with reservoir frequency $\omega_l$ and chemical potential $\mu_{\alpha_l}$. The reservoir contraction reads
\beq
  \gamma_{21}^{p_2p_1}=\delta_{2\overline{1}} p_1 f(-\xi_1 p_1 \omega_1)
\eeq
with Fermi function $f$, and index notation `$\overline{1}$'$=(-\xi_1,k_1,\alpha_1)$. Finally, the sign forefactors in \eq{sig2translate} and \eq{sig4Xtranslate} come from a product of $(-i)^n$ for the $n$th order diagram and Wick sign $(-1)^{N_p}$ with $N_p$ the permutation number of the digram (here $N_p=0$ for the sequential term and  $N_p=1$ for the exchange).
In all diagrams, summations over all $\xi$, $\alpha$, and $p$, as well as integrations over all bath frequencies, are implied.
As in the diagrams of \cite{lei08}, one could label each tunnel vertex with appropriates indices, and each propagator with appropriate frequency index. However, since these are given uniquely by position in diagram, there  is little advantage in doing so.  
We will, however, label the propagators with Laplace argument $z$, and/or $\chi$ when appropriate, as our formal manipulations require explicit functions of these variables.

The self-consistent equations of Eqs. (\ref{SIGa}), (\ref{SIGb}) and (\ref{SIGc}) include diagrams as above but with propagator
\beq
  \underset{z}{\eqcirc} \to \frac{1}{z_m - {\cal L}_\mathrm{S} -\Sigma(z)}
  ,
\eeq
where the self-energy $\Sigma(z)$ is the appropriate one for the calculation in hand, e.g. $\Sigma^{(a)}$, etc..
Moreover, for calculating the counting statistics, the relevant propagators are the $\chi$-dependent ones
\beq
  \underset{\chi;z}{\eqcirc}
  \to 
  \frac{1}{z_m - {\cal L}_\mathrm{S} -\Sigma(z;\chi)}
  .
\eeq
In discussing $\chi$-dependent quantities, multiplying a diagram by
counting-field factor $q(\chi)$, corresponds to multiplication of the corresponding super-operator by the same quantity (under the summation ofcourse).
For example, the full $z$ and $\chi$ dependent sequential self-energy diagram of \eq{SIGchia} reads
\beq
  \Sigma^{(a)}(z;\chi) &=& 
  \contraction{}{G}{\underset{z;\chi}{\eqcirc}}{G}
  G \underset{z;\chi}{\eqcirc} G 
  \;q_2(\chi)
  ,
\eeq
and this translates as
\beq
  \Sigma^{(a)}(\chi;z) &=& 
  G^{p_2}_{\bar{1}} 
  \frac{- p_1 f(-\xi_1 p_1 \omega_1)
  e^{
     i s_{\alpha_1} \xi_1 \frac{1}{2}\rb{p_1-p_2}\chi_{\alpha_1}
  }}
  {z-x_1 
    -{\cal L}_\mathrm{S}-\Sigma^{(a)}(z;\chi)
  } 
  G^{p_1}_1
  ,
  \nonumber
\eeq
where we have used the explicit form of $q_2(\chi)$ from \eq{q2}.

\subsection{Kernel expansions}
Insight into the approximations provided by our self-consistent kernels can be obtained by expanding the effective propagators
\beq
  \rb{\underset{z}{\eqcirc}}  = 
  \rb{\underset{z}{-}}
  +
  \rb{
    \underset{z}{-}\Sigma(z)\underset{z}{-}
  }
  +
  \rb{
    \underset{z}{-}\Sigma(z)\underset{z}{-}\Sigma(z)\underset{z}{-}
  }
  +
  \ldots
  \nonumber
  .
\eeq
In this way, we can expand the SC self-energy $\Sigma^{(a)}(z)$ of \eq{SIGa} as
\beq
  \Sigma^{(a)}(z) = 
  \contraction{}{G}{\underset{z-x_1}{-}}{G}
  G \underset{z-x_1}{-} G
  +
  \contraction[2ex]{}{G} {\underset{z-x_1}{-} G \underset{z-x_2}{-} G \underset{z-x_1}{-}} {G}
  \contraction{G \underset{z-x_2 }{-}} {G} {\underset{z-x_1}{-}} {G} 
  G \underset{z-x_1}{-} G \underset{z-x_2}{-} G \underset{z-x_1}{-} G
  +\ldots
  ,
  \label{expansion}
  \nonumber\\
\eeq 
where we have made explicit the $x$ variables in the propagators and, in doing so, have taken into account the $\delta$-functions in the bath contractions.  The expansion of  $\Sigma^{(a)}(z) $ therefore generates a diagram with the same topology as the direct cotunneling term.  With $x$-arguments, the exact direct diagram reads
\beq
  \Sigma^{(4D)}(z) =  
  \contraction[2ex]{}{G} {\underset{z-x_1}{-} G \underset{z-x_1-x_2}{-} G \underset{z-x_1}{-}} {G}
  \contraction{G \underset{z-x_1 }{-}} {G} {\underset{z-x_1-x_2}{-}} {G} 
  G \underset{z-x_1}{-} G \underset{z-x_1-x_2}{-} G \underset{z-x_1}{-} G
  \label{DirectX}
  .
\eeq
Comparison of \eq{expansion} and \eq{DirectX} shows that the fourth-order term in the SC expansion provides an approximation which differs only the $x$ argument of the middle propagator.

\section{Matrix representation\label{appMATRIX}}

In this Liouville-space approach, superoperators such as the Liouvillian can be represented as finite-dimensional matrices and we often make use here of their eigendecomposition.
Of particular importance is the decomposition of the $\chi=0$, $z=0$ effective Liouvillian ${\cal W}= {\cal W}(0,0^+)$, from which the stationary properties of the system may be calculated.
Since ${\cal W}$ is nonHermitian, it has non-adjoint left, $\braa{\psi_a}$,
and right, $\kett{\psi_a}$, eigenvectors defined by the equations 
\beq
  {\cal W} \kett{\psi_a}  &=& w_a \kett{\psi_a}
  \nonumber\\
  \braa{\psi_a} {\cal W}   &=& w_a \braa{\psi_a},
\eeq
with eigenvalues $w_a$ and index $a=0,1,\ldots,N-1$ with $N$ the dimension of ${\cal W}$.
Taken together the left and right eigenvectors form a bi-orthonormal set:
$
  \eww{\psi_a|\psi_{a'}} = \delta_{a,a'}
$ and we have the closure relation in Liouvillian space
\beq
  \mathbbm{1} = \sum_{a}\opp{\psi_a}{\psi_a} 
  \label{cset}
  .
\eeq
The eigendecomposition of the stationary kernel reads
\beq
  {\cal W} = \sum_{a}w_a \opp{\psi_a}{\psi_a} 
  \label{eigenD}
  .
\eeq
Assuming that the system has a unique stationary state, one (and only one) of the eigenvalues will be zero, $w_0 =0$, and the
stationary state itself is given by the corresponding right eigenvector $\rho^\mathrm{stat}_\mathrm{S} = \kett{\psi_0}$.  The dual vector $ \braa{\psi_0}$ is then the ``trace vector'' with elements unity at positions mapping to populations and zero at those mapping to coherences.  The remaining eigenvalues of ${\cal W}$ have non-positive real parts and, if complex, occur in complex conjugate pairs.

\section{Integrals \label{appINTEGRALS}}

We now discuss the analytic evaluation of the integrals required here.  For convenience we set $k_BT=1$.
The $z=0$ integral equation for $\Sigma^{(a)}(z=0)$ reads
\beq
  \Sigma^{(a)}(0) &=& 
  \int d \omega_1
  G^{p_2}_{\bar{1}} 
  \frac{- p_1 f(-\xi_1 p_1 \omega_1)}
  {0^++i\xi_1(\omega_1+\mu_{\alpha_1})
    -{\cal W}^{(a)}(0)
  } 
  G^{p_1}_1
  \nonumber
  .
\eeq
Using the eigendecomposition for ${\cal W}^{(a)}(0)$, as in \eq{eigenD}, we write this as
\beq
   \Sigma^{(a)}(0) &=& 
  - p_1 G^{p_2}_{\bar{1}} 
   \opp{\psi_a}{\psi_a} 
  G^{p_1}_1
  \nonumber\\
  &&
  \times \int d \omega_1
  \frac{ f(-\xi_1 p_1 \omega_1)}
  {0^++i\xi_1(\omega_1+\mu_{\alpha_1})
    -w_a
  }\nonumber\\
  &=&
  - 2\pi p_1 G^{p_2}_{\bar{1}} 
   \opp{\psi_a}{\psi_a} 
  G^{p_1}_1
  ~
  I^{(2)}_{p_1}(\xi_1\mu_{\alpha_1} + i w_a)
  \nonumber
  ,
\eeq
which defines the sequential integral, $I_{p}^{(2)}(\lambda)$.  This integral can be evaluated as in I with the introduction of a Lorentzian cut-off function 
$
  {\cal D}(\omega)= X_C^2/\left\{(\omega-\omega_0)^2 + X_C^2\right\}
$,
with band-centre $\omega_0$ chosen for convenience,
as
\beq
  I_{p}^{(2)}(\lambda) 
   \equiv\frac{i}{2\pi} 
	\int d \omega
	\frac{f(\omega)}{i0^+ + p \omega - \lambda}
  &=&  
  \frac{1}{2} f( p\lambda)  +\frac{i p}{2\pi}\phi(\lambda)
  \nonumber
  .
\eeq
Here
$
  \phi (\omega) = \frac{1}{2}
  \rb{\frac{}{}
  g(\omega)+ g(-\omega)
  -g(\omega+i X_c)-g(-\omega+i X_c)
  }
  \nonumber
$,
with
$
  g(\omega) \equiv \Psi\rb{\frac{1}{2} + \frac{\omega}{2\pi iT}}
$,
and $\Psi$ the digamma function.

\begin{widetext}
As discussed in I, evaluation of fourth-order terms requires the integral
\beq
    I^{p_1 p_2}(\lambda_2, \lambda_3) 
    &=&
    \int d\omega_1 d\omega_2
    \frac{f(p_1 \omega_1)f(p_2 \omega_2)}{(i0^+ +\omega_1+\omega_2-\lambda_2)(i0^+ + \omega_1 -\lambda_3)}
    .
\eeq
In straightforward Liouvillian Perturbation theory, the $\lambda_i$ are real; for self-consistent calculations, they can be complex and this means that we need the result for this integral in a slightly more general form than given in I.  We have
\beq
     I^{p_1 p_2}(\lambda_2, \lambda_3)
    &=&
    p_1 p_2 G(\lambda_2, \lambda_3) + p_1 \widetilde{G}(\lambda_3)
    + p_1 p_2 H(\lambda_2,\lambda_3) + \mathrm{const}
  \eeq
  with ``const'' an unimportant constant.  The first two part functions are
  \beq
    G(\lambda_2, \lambda_3) &=&
      2 \pi^2 (\textstyle{\frac{1}{2}}+b(\lambda_2)) 
      \left\{ I^{(2)}_{+1}(\lambda_3)
      - I^{(2)}_{+1}(\lambda_{32})\right\}
      +g(-\lambda_3) \phi(\lambda_{32})
    \nonumber\\
    &&
      -\textstyle{\frac{1}{2}} g(-\lambda_3+ i X_C) \phi(\lambda_{32}-i X_C)
      -\textstyle{\frac{1}{2}} g(\lambda_3+ i X_C) \phi(\lambda_{32}+i X_C)
      ,
  \eeq
  and
  \beq
    \widetilde{G}(\lambda_3)=-\pi^2 I^{(2)}_{+1}(\lambda_3),
  \eeq
with $\lambda_{ij}=\lambda_i-\lambda_j$ and Bose-Einstein distribution function $b(x) = \rb{e^{x}-1}^{-1}$. The third contribution is more complicated and reads
  \beq
   H(\lambda_2,\lambda_3) &=&
   \sum_{k=1,3,5,\ldots}^\infty
     g(-\lambda_2 + i \pi k)
     \left\{
	   \frac{1}{\lambda_{32}-iX_C +i\pi k}
	   + \frac{1}{\lambda_{32}+iX_C+i\pi k}
	   - \frac{2}{\lambda_{32}+i\pi k}
     \right\}
	 \nonumber\\
	 && ~~~~~~
	 -      
     g(-\lambda_2 + i X_C + i \pi k)
	 \left\{
	   \frac{1}{\lambda_{32}+i\pi k}
	    + \frac{1}{\lambda_{32}+2iX_C+i\pi k}
	    - \frac{2}{\lambda_{32}+iX_C+i\pi k}
	 \right\}
	 \nonumber\\
	 && ~~~~~~
	 +g(\lambda_2 + i \pi k)
     \left\{
	   \frac{1}{-\lambda_{32}-iX_C +i\pi k}
	   + \frac{1}{-\lambda_{32}+iX_C+i\pi k}
	   - \frac{2}{-\lambda_{32}+i\pi k}
     \right\}
	 \nonumber\\
	 && ~~~~~~
	 -      
     g(\lambda_2 + i X_C + i \pi k)
	 \left\{
	   \frac{1}{-\lambda_{32}+i\pi k}
	    + \frac{1}{-\lambda_{32}+2iX_C+i\pi k}
	    - \frac{2}{-\lambda_{32}+iX_C+i\pi k}
	 \right\}
	 .
  \eeq
It remains to discuss the translation and integration of diagrams of the form
$
   \contraction{}{G}{\eqcirc M\eqcirc}{G}
  G \eqcirc M \eqcirc G 
$,
such as occur in \eq{J'} for example.   The analytic expression of this diagram reads
\beq
   - p_1 \int d \omega_1
  G^{p_2}_{\bar{1}} 
  \frac{1}
  {0^++i\xi_1(\omega_1+\mu_{\alpha_1})
    -{\cal W}^{(a)}(0)
  } 
  M
  \frac{1}
  {0^++i\xi_1(\omega_1+\mu_{\alpha_1})
    -{\cal W}^{(a)}(0)
  } 
  G^{p_1}_1
  f(-\xi_1 p_1 \omega_1)
\eeq 
Introduction of the eigendecomposition for the stationary kernels and the 
employ of a partial fraction decomposition allows us to write this as
\beq
  - p_1 G^{p_2}_{\bar{1}} 
   \opp{\psi_a}{\psi_a}  
   M
   \opp{\psi_{a'}}{\psi_{a'}} 
  G^{p_1}_1
  ~
  \int d \omega_1 
  \frac{f(-\xi_1 p_1 \omega_1)}{w_a-w_{a'}}
  \rb{
	  \frac{1}
	  {0^++ i\xi_1(\omega_1+\mu_{\alpha_1})
	    -w_a
	  }
	  -
	   \frac{1}
	  {0^++ i\xi_1(\omega_1+\mu_{\alpha_1})
	    -w_{a'}
	  }
  }
  \nonumber
\eeq
Identifying the sequential integrals, this becomes
\beq
   - 2 \pi p_1 G^{p_2}_{\bar{1}} 
   \opp{\psi_a}{\psi_a}  
   M
   \opp{\psi_{a'}}{\psi_{a'}} 
  G^{p_1}_1
  ~
  \frac{
  I^{(2)}_{p_1}(\mu_{\alpha_1} + i w_a)
  -
  I^{(2)}_{p_1}(\mu_{\alpha_1} + i w_{a'})
  }{w_a-w_{a'}}
  .
\eeq
In the case that $w_a=w_{a'}$, the differential quotient is recognised and  the integral part replaced with
$
d/dw_a I^{(2)}_{p_1}(\mu_{\alpha_1} + i w_a)
$.
The evaluation of similar cotunneling-type terms follows analogously.

\section{Jump-operator equations for the $\Sigma^{(b)}$ SC kernel \label{appSIGB}}

The four equations corresponding to Eqs (\ref{J'}-\ref{J'dot}) for the current blocks of $\Sigma^{(b)}$ are:
\beq
  {\cal J}' &=& 
  \contraction{}{G}{\eqcirc}{G}
  G \eqcirc G ~q_2'
  +
  \contraction {}{G} {\eqcirc G \eqcirc} {G}
  \contraction[2ex]{G\eqcirc} {G} {\eqcirc G \eqcirc} {G}
  G \eqcirc G \eqcirc G\eqcirc G ~ q_{4X}'
  +
  \contraction{}{G}{\eqcirc{\cal J}'\eqcirc}{G}
  G \eqcirc{\cal J}'\eqcirc G 
  +
  \contraction {}{G} {\eqcirc {\cal J}'\eqcirc G\eqcirc} {G}
  \contraction[2ex]{G\eqcirc{\cal J}'\eqcirc} {G} {\eqcirc G \eqcirc} {G}
  G \eqcirc {\cal J}'\eqcirc G \eqcirc G\eqcirc G 
  \nonumber\\
  &&
  +~
  \contraction {}{G} {\eqcirc G \eqcirc{\cal J}'\eqcirc} {G}
  \contraction[2ex]{G\eqcirc} {G} {\eqcirc {\cal J}'\eqcirc G \eqcirc} {G}
  G \eqcirc G \eqcirc{\cal J}'\eqcirc G\eqcirc G
  +
  \contraction {}{G} {\eqcirc G\eqcirc} {G}
  \contraction[2ex]{G\eqcirc} {G} {\eqcirc G \eqcirc{\cal J}'\eqcirc} {G}
  G \eqcirc G \eqcirc G\eqcirc {\cal J}'\eqcirc G 
\eeq

\beq
  \dot{{\cal J}} &=& 
  \contraction{}{G}{\eqcirc\rb{i\mathbbm{1}+\dot{{\cal J}}}\eqcirc}{G}
  G \eqcirc\rb{i \mathbbm{1}+\dot{{\cal J}}}\eqcirc G 
  +
  \contraction {}{G} {\eqcirc \rb{i\mathbbm{1}+\dot{{\cal J}}}\eqcirc G\eqcirc} {G}
  \contraction[2ex]{G\eqcirc\rb{i\mathbbm{1}+\dot{{\cal J}}}\eqcirc} {G} {\eqcirc G \eqcirc} {G}
  G \eqcirc\rb{i\mathbbm{1}+\dot{{\cal J}}}\eqcirc G \eqcirc G\eqcirc G 
  \nonumber\\
  &&
  +~
  \contraction {}{G} {\eqcirc G \eqcirc\rb{i\mathbbm{1}+\dot{{\cal J}}}\eqcirc} {G}
  \contraction[2ex]{G\eqcirc} {G} {\eqcirc\rb{i\mathbbm{1}+\dot{{\cal J}}}\eqcirc G \eqcirc} {G}
  G \eqcirc G \eqcirc\rb{i\mathbbm{1}+\dot{{\cal J}}}\eqcirc G\eqcirc G
  +
  \contraction {}{G} {\eqcirc G\eqcirc} {G}
  \contraction[2ex]{G\eqcirc} {G} {\eqcirc G \eqcirc\rb{i\mathbbm{1}+\dot{{\cal J}}}\eqcirc} {G}
  G \eqcirc G \eqcirc G\eqcirc \rb{i\mathbbm{1}+\dot{{\cal J}}}\eqcirc G 
\eeq

\beq
  \eww{{\cal J}''} &=& 
  \eww{
	  \contraction{}{G}{\eqcirc}{G}
	  G \eqcirc G ~q_2''
	  +
	  \contraction {}{G} {\eqcirc G \eqcirc} {G}
	  \contraction[2ex]{G\eqcirc} {G} {\eqcirc G \eqcirc} {G}
	  G \eqcirc G \eqcirc G\eqcirc G ~ q_{4X}''
	}
  \nonumber\\
  &&
  +~
  2 
  \eww{  
	  \contraction{}{G}{\eqcirc{\cal J}'\eqcirc}{G}
	  G \eqcirc{\cal J}'\eqcirc G ~q_2'
	  +  
	  \contraction {}{G} {\eqcirc {\cal J}'\eqcirc G\eqcirc} {G}
	  \contraction[2ex]{G\eqcirc{\cal J}'\eqcirc} {G} {\eqcirc G \eqcirc} {G}
	  G \eqcirc {\cal J}'\eqcirc G \eqcirc G\eqcirc G  ~q_{4X}'
    +
	  \contraction {}{G} {\eqcirc G \eqcirc{\cal J}'\eqcirc} {G}
	  \contraction[2ex]{G\eqcirc} {G} {\eqcirc {\cal J}'\eqcirc G \eqcirc} {G}
	  G \eqcirc G \eqcirc{\cal J}'\eqcirc G\eqcirc G  ~q_{4X}'
	  +
	  \contraction {}{G} {\eqcirc G\eqcirc} {G}
	  \contraction[2ex]{G\eqcirc} {G} {\eqcirc G \eqcirc{\cal J}'\eqcirc} {G}
	  G \eqcirc G \eqcirc G\eqcirc {\cal J}'\eqcirc G   ~q_{4X}'
	}
	\nonumber\\
\eeq

\beq
  \eww{\dot{{\cal J}}'} &=& 
  \eww{
	  \contraction{}{G}{\eqcirc\rb{i\mathbbm{1}+\dot{{\cal J}}}\eqcirc}{G}
	  G \eqcirc\rb{i \mathbbm{1}+\dot{{\cal J}}}\eqcirc G ~q_2'
	  +
	  \contraction {}{G} {\eqcirc \rb{i\mathbbm{1}+\dot{{\cal J}}}\eqcirc G\eqcirc} {G}
	  \contraction[2ex]{G\eqcirc\rb{i\mathbbm{1}+\dot{{\cal J}}}\eqcirc} {G} {\eqcirc G \eqcirc} {G}
	  G \eqcirc\rb{i\mathbbm{1}+\dot{{\cal J}}}\eqcirc G \eqcirc G\eqcirc G ~q_{4X}'
	}
  \nonumber\\
  &&
  +~
  \eww{
	  \contraction {}{G} {\eqcirc G \eqcirc\rb{i\mathbbm{1}+\dot{{\cal J}}}\eqcirc} {G}
	  \contraction[2ex]{G\eqcirc} {G} {\eqcirc\rb{i\mathbbm{1}+\dot{{\cal J}}}\eqcirc G \eqcirc} {G}
	  G \eqcirc G \eqcirc\rb{i\mathbbm{1}+\dot{{\cal J}}}\eqcirc G\eqcirc G~q_{4X}'
	  +
	  \contraction {}{G} {\eqcirc G\eqcirc} {G}
	  \contraction[2ex]{G\eqcirc} {G} {\eqcirc G \eqcirc\rb{i\mathbbm{1}+\dot{{\cal J}}}\eqcirc} {G}
	  G \eqcirc G \eqcirc G\eqcirc \rb{i\mathbbm{1}+\dot{{\cal J}}}\eqcirc G~q_{4X}'
	}
	.
\eeq
The corresponding blocks for $\Sigma^{(c)}$ follow analogously.
\end{widetext}




\begin{thebibliography}{99}
\bibitem{leftmostG}  Diagrams which have an undifferentiated counting-field function (e.g. $q_2$) have as leftmost vertex a plain tunnel vertex, $G$, since current vertices are obtained when counting-field functions are differentiated with respect to $\chi$.  Such diagrams disappear in the expectation value since the sum over the Keldysh index of a leftmost tunnel vertex yields two contributions which, under the trace, are identical but with opposite sign and these contributions thus cancel.  A particular example is the expectation value of any one of the self-energies $\eww{\Sigma(\chi=0)}=0$ since $\Sigma(\chi=0)$ contains only tunnel vertices.
\end{thebibliography}
\end{document}